# The Dynamics of a Modified Jaynes-Cummings Model

Moorad Alexanian

*Department of Physics and Physical Oceanography*
*University of North Carolina Wilmington, Wilmington, NC 28403-5606*

Email: alexanian@uncw.edu

**Abstract.** We introduce a dynamical system that instead of exchanging a single photon as in the atomic system of the usual Jaynes-Cummings model (JCM), it exchanges instead a squeezed coherent photon. Accordingly, the creation and annihilation photon operators in the JCM are replaced by the creation and annihilation operators of squeezed coherent states, respectively. This transformation generates a photon Hamiltonian that modifies the JCM that includes terms usually omitted in the JCM when making the rotating-wave approximation and, in addition, the transformation introduces atom exchange terms absent in the JCM.

## 1. Introduction

The Jaynes-Cummings model (JCM) [1] of a two-level atomic system coupled to a single-mode radiation field is known to exhibit interesting optical phenomena, such as the collapse and revival of Rabi oscillations of the atomic coherence [2–5]. There have been may attempts to extend the model to deal with differing physical phenomena. A generalized kinematic approach to compute the geometric phases acquired in both unitary and dissipative Jaynes-Cummings models, which provide a fully quantum description for a two-level system interacting with a single mode of the (cavity) electromagnetic field, in a perfect or dissipative cavity, respectively [6]. The calculation of the cavity- field distribution in the Wigner representation for the two-photon resonance of the weakly driven Jaynes-Cummings oscillator in its strong-coupling limit. Using an effective four-level system, one analytically demonstrate the presence of steady state and transient bimodality, which breaks azimuthal symmetry in phase space [7]. The effective simulation of light-matter ultrastrong-coupling phenomena with strong-coupling systems. Recent theory and experiments have shown that the quantum Rabi Hamiltonian can be simulated by a Jaynes-Cummings system with the addition of two classical drives. This allows one to implement nonlinear processes that do not conserve the total number of excitations [8]. A reliable scheme to realize the ultrastrong Jaynes-Cummings model is proposed by simultaneously modulating the resonance frequencies of the two-level system and the bosonic mode in the ultrastrong quantum Rabi model [9].

In this paper, we introduce a dynamical system that instead of exchanging a single photon as in the atomic system of the usual JCM, it exchanges instead a squeezed coherent photon. Accordingly, the creation and annihilation photon operators $\hat{a}^\dagger$ and $\hat{a}$ in the JCM are replaced by the creation and annihilation operators of squeezed coherent states, viz., $\hat{B}^\dagger$ and $\hat{B}$, respectively. This transformation generates a photon Hamiltonian that modifies the JCM that includes terms usually omitted in the JCM when making the rotating-wave approximation and, in addition, the transformation introduces atom exchange terms absent in the JCM. This paper is arranged as follows. In Sec. 2, we review the creation and annihilation operators for squeezed coherent states. In Sec. 3, we introduce the generalized JCM. In Sec. 4, we calculate the dynamical behavior of our system for an initial photon coherent state. In Sec. 5, we study the collapse and revival of our dynamically generated state. Finally, Sec. 6 summarizes our results.

## 2. Squeezed coherent photons

In recent paper [10], the creation and annihilation operators $\hat{B}$ and $\hat{B}^\dagger$, respectively, for the squeezed coherent photons are given by, where $\hat{a}$ and $\hat{a}^\dagger$ are the photon creation and annihilation operators,

$$\hat{B} = \hat{S}(\zeta)\hat{D}(\alpha)\hat{a}\hat{D}(-\alpha)\hat{S}(-\zeta) = \cosh(r)\hat{a} + e^{i\varphi}\sinh(r)\hat{a}^\dagger - \alpha \tag{1}$$

and

$$\hat{B}^\dagger = \hat{S}(\zeta)\hat{D}(\alpha)\hat{a}^\dagger\hat{D}(-\alpha)\hat{S}(-\zeta) = e^{-i\varphi}\sinh(r)\hat{a} + \cosh(r)\hat{a}^\dagger - \alpha^*, \tag{2}$$

with inverses

$$\hat{a} = \hat{D}(-\alpha)\hat{S}(-\zeta)\hat{B}\hat{S}(\zeta)\hat{D}(\alpha) = \cosh(r)\hat{B} - e^{i\varphi}\sinh(r)\hat{B}^\dagger + \alpha\cosh(r) - \alpha^* e^{i\varphi}\sinh(r) \tag{3}$$

and

$$\hat{a}^\dagger = \hat{D}(-\alpha)\hat{S}(-\zeta)\hat{B}^\dagger\hat{S}(\zeta)\hat{D}(\alpha) = -e^{-i\varphi}\sinh(r)\hat{B} + \cosh(r)\hat{B}^\dagger + \alpha^*\cosh(r) - \alpha e^{-i\varphi}\sinh(r), \tag{4}$$

where

$$\hat{D}(\alpha) = \exp(\alpha\hat{a}^\dagger - \alpha^*\hat{a}) \tag{5}$$

is the Glauber displacement operator with $\alpha = |\alpha|\exp(i\theta)$ and

$$\hat{S}(\zeta) = \exp\left(-\frac{\zeta}{2}\hat{a}^{\dagger 2} + \frac{\zeta^*}{2}\hat{a}^2\right) \tag{6}$$

is the squeezing operator with $\zeta = r\exp(i\varphi)$. Note that $[\hat{B}, \hat{B}^\dagger] = 1$ follows from $[\hat{a}, \hat{a}^\dagger] = 1$.

One can generate a squeezed coherent state from the vacuum as follows

$$|\zeta, \alpha, 0\rangle \equiv \hat{S}(\zeta)\hat{D}(\alpha)|0\rangle, \tag{7}$$

where the state $|\zeta, \alpha, 0\rangle$ is the vacuum state since $B|\zeta, \alpha, 0\rangle = 0$. Consider the state

$$\frac{(\hat{B}^\dagger)^n}{\sqrt{n!}}|\zeta, \alpha, 0\rangle = \hat{S}(\zeta)\hat{D}(\alpha)\frac{(\hat{a}^\dagger)^n}{\sqrt{n!}}|0\rangle \equiv |\zeta, \alpha, n\rangle, \tag{8}$$

where the state $|\zeta, \alpha, n\rangle$ is called an *n*-photon squeezed coherent state with the ordinary squeezed coherent state $|\zeta, \alpha, 0\rangle$ being the 0-photon coherent state. It follows from (1)-(4) and (8) that

$$\begin{aligned}\hat{B}|\zeta, \alpha, n\rangle &= \sqrt{n}|\zeta, \alpha, n-1\rangle \\ \hat{B}^\dagger|\zeta, \alpha, n\rangle &= \sqrt{n+1}|\zeta, \alpha, n+1\rangle \\ \hat{B}^\dagger\hat{B}|\zeta, \alpha, n\rangle &= n|\zeta, \alpha, n\rangle.\end{aligned} \tag{9}$$

## 3. Generalized Jaynes-Cummings model

Consider the modified JCM

$$\hat{H}_I = \frac{\hbar\Delta}{2}\hat{\sigma}_3 - i\hbar\lambda(\hat{\sigma}_+\hat{B} - \hat{B}^\dagger\hat{\sigma}_-), \tag{10}$$

where $\hat{N} = \hat{B}^\dagger\hat{B} + \hat{\sigma}_+\hat{\sigma}_-$ is a constant of the motion and represents the number of quanta and the detuning $\Delta = \omega_2 - \omega_1 - \omega$, with $\hbar\omega_1$, $\hbar\omega_2$ are the energies of the uncoupled states $|1\rangle$ and $|2\rangle$, respectively, and ω is the frequency of the field mode. The system can be in two possible states $|i\rangle$, $i = 1, 2$ with $|1\rangle$ being the ground state of the system and $|2\rangle$ being the excited state, respectively. The transition in the JCM is based on the exchange of one photon, here the transition is via a squeezed coherent photon. The four Paul operators are

$$\begin{aligned}1 &= |2\rangle\langle 2| + |1\rangle\langle 1| \\ \hat{\sigma}_3 &= |2\rangle\langle 2| - |1\rangle\langle 1| \\ \hat{\sigma}_+ &= |2\rangle\langle 1| \\ \hat{\sigma}_- &= |1\rangle\langle 2|.\end{aligned} \qquad (11)$$

The Schrödinger equation of motion is

$$i\hbar \frac{d}{dt}|\psi_I(t)\rangle = \hat{H}_I |\psi_I(t)\rangle \qquad (12)$$

with

$$|\psi_I(t)\rangle = \sum_{n=0}^{\infty} [c_{1,n}(t)|1\rangle + c_{2,n}(t)|2\rangle]|\zeta, \alpha, n\rangle \qquad (13)$$

and so

$$\begin{aligned}\dot{c}_{1,n}(t) &= \frac{i\Delta}{2} c_{1,n}(t) + \lambda n^{1/2} c_{2,n-1}(t) \\ \dot{c}_{2,n-1}(t) &= -\frac{i\Delta}{2} c_{2,n-1}(t) - \lambda n^{1/2} c_{1,n}(t).\end{aligned} \qquad (14)$$

for the probability amplitudes with initial state

$$|\psi_I(0)\rangle = \sum_{n=0}^{\infty} b_n |\zeta, \alpha, n\rangle |1\rangle. \qquad (15)$$

These equations are precisely the same as in the ordinary JCM with solutions [13]

$$\begin{aligned}c_{1,n}(t) &= b_n \left( \cos[\Omega_R(n) t/2] + i\frac{\Delta}{\Omega_R(n)} \sin[\Omega_R(n) t/2] \right) \\ c_{2,n-1}(t) &= -b_n \frac{2\lambda n^{1/2}}{\Omega_R(n)} \sin[\Omega_R(n) t/2],\end{aligned} \qquad (16)$$

where $\Omega_R(n) = \sqrt{\Delta^2 + 4\lambda^2 n}$ is the photon-number dependent Rabi frequency.

The Hamiltonian (10), when expressed in terms of the photon operators via (1) and (2), gives rise to terms neglected when deriving the JCM that is based on the rotating-wave approximation

$$\hat{H}_I = \frac{\hbar\Delta}{2}\hat{\sigma}_3 + i\hbar\lambda(\alpha\hat{\sigma}_+ - \alpha^*\hat{\sigma}_-) - i\hbar\lambda\cosh(r)(\hat{\sigma}_+\hat{a} - \hat{a}^\dagger\hat{\sigma}_-) - i\hbar\lambda\sinh(r)(e^{i\varphi}\hat{\sigma}_+\hat{a}^\dagger - e^{-i\varphi}\hat{\sigma}_-\hat{a}). \qquad (17)$$

The last term in (17) gives rise to corrections to the rotating-wave approximation. In fact, for $\varphi = \pi$, $\alpha = 0$, and $r \gg 1$ such that, $\sinh(r) \approx \cosh(r) \approx e^r/2$ one has that

$$\hat{H}_I \approx \frac{\hbar\Delta}{2}\hat{\sigma}_3 - i\hbar\lambda\frac{e^r}{2}(\hat{\sigma}_+ + \hat{\sigma}_-)(\hat{a} - \hat{a}^\dagger), \qquad (18)$$

which is the Hamiltonian prior to making the rotating-wave approximation, albeit, in the ultrastrong coupling limit since $\lambda \to \lambda e^r/2$.

## 4. Coherent initial state

Consider the initial state with the radiation field in a coherent state and the atom in its ground state, viz.,

$$|\psi_I(0)\rangle = \hat{D}(\beta)|0\rangle|1\rangle. \tag{19}$$

One obtains, with the aid of (8) and (15), that

$$b_n = \langle n|\hat{D}(-\alpha)\hat{S}(-\zeta)\hat{D}(\beta)|0\rangle. \tag{20}$$

Now

$$\hat{S}(-\zeta)\hat{D}(\beta) = \hat{D}(\gamma)\hat{S}(-\zeta), \tag{21}$$

where $\gamma = \beta \cosh(r) + \beta^* e^{i\varphi} \sinh(r)$. Also,

$$\hat{D}(-\alpha)\hat{D}(\gamma) = e^{-(\alpha\gamma^* - \gamma\alpha^*)/2}\hat{D}(\gamma - \alpha). \tag{22}$$

Therefore,

$$b(n) = \sqrt{\frac{\text{sech}(r)}{n!}}(\gamma - \alpha)^n e^{(\gamma\alpha^* - \alpha\gamma^* - |\gamma - \alpha|^2)/2} \sum_{l=0}^{\infty} \frac{1}{2^l l!}(\alpha^* - \gamma^*)^{2l}(e^{i\varphi}\tanh(r))^l {}_0F_2(0; -2l; -n, -1/|\gamma - \alpha|^2), \tag{23}$$

where the generalized hypergeometric series [12]

$${}_0F_2(0; -2l, -n; -1/|\gamma - \alpha|^2) = \sum_{j=0}^{2l} \frac{\Gamma(2l+1)\Gamma(n+1)}{\Gamma(2l+1-j)\Gamma(n+1-j)} \frac{(-1/|\gamma - \alpha|^2)^j}{j!}, \tag{24}$$

and the infinity series reduces to a polynomial since $l$ is a positive integer.

In (23) we choose the phases of $\alpha = ae^{i\theta}$, $\zeta = re^{i\varphi}$, and $\beta = be^{i\chi}$ as follows: $\varphi = 2\theta = 2\chi$ and so (23) becomes,

$$b(n, a, b, r) = \sqrt{\frac{\text{sech}(r)}{n!}}(be^r - a)^n e^{-(be^r - a)^2/2} e^{i\chi n} \sum_{l=0}^{\infty} \frac{1}{2^l l!}(be^r - a)^{2l}\tanh^l(r) \, {}_0F_2(0; -2l; -n, -1/(be^r - a)^2). \tag{25}$$

## 5. Collapse and revival

In order to contrast the dynamics generated by the Hamiltonian (17) and that generated by the JCM, obtained from (17) with $r = 0$ and $\alpha = 0$, consider the behavior of the corresponding collapse and revival for the initial state (19). We are interested in the probability of the system being in the ground state $|1\rangle$ at time $t$ for the resonant case ($\Delta = 0$) and so

$$P(t) = \sum_{n=0}^{\infty} |c_{1,n}(t)|^2 = \frac{1}{2} \sum_{n=0}^{\infty} |b_n|^2 [1 + \cos(2\lambda\sqrt{n}t)]. \tag{26}$$

while for the JCM we have

$$P_{JCM}(t) = \frac{1}{2} \sum_{n=0}^{\infty} \frac{e^{-|\beta|^2}|\beta|^{2n}}{n!}[1 + \cos(2\lambda\sqrt{n}t)]. \tag{27}$$

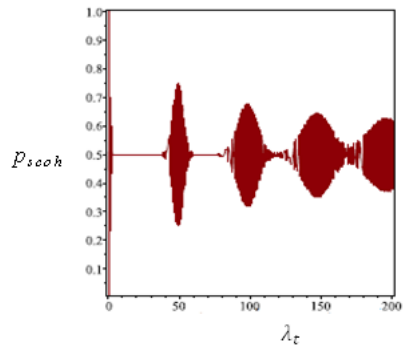

**Fig. 1.** Plot for the probability $p_{scoh}(t)$ for the collapse and revival given in Eq. (26) with $|\alpha| = 10$, $|\beta| = 2$, and $|\zeta| = 0.1$ in (25), that is, with an initial coherent state with four photons.

In Fig. 1 we see the contrast of the collapse and revival of the model given by (17) with $|\alpha| = 10$ and $|\zeta| = 0.1$ for the initial coherent state with $|\beta| = 2$ and of the behavior for coherent state also with $\beta = 2$ shown in Fig. 2. Notice the drastic difference. In Fig. 3 we show the effect of the presence of the fourth term in (17) in the absence of the second term on (17) in the collapse and revival with $|\alpha| = 0$, $|\beta| = 5$ and $|\zeta| = 0.9$. This elucidates the effect of the terms that are neglected in the rotating-wave approximation in obtaining the JCM. Fig. 4 shows the behavior of the coherent state given by (27) for $|\beta| = 5$, that is, with twenty-five initial photons. Finally, in Fig 5, we do the converse and see the effect of the presence of the second term in (17) in the absence of the fourth term in (17). Fig. 5 has $|\alpha| = 15$, $|\beta| = 5$, and $|\zeta| = 0$ to be compared to the behavior of the coherent state of Fig. 4 also with $|\beta| = 5$.

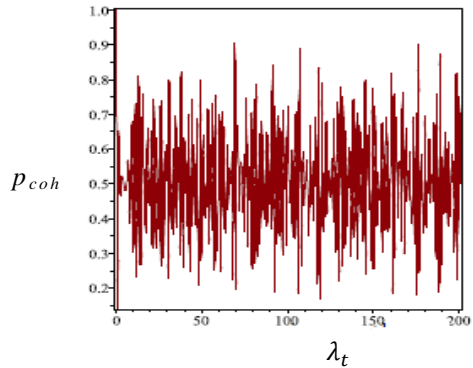

**Fig. 2.** Plot for the probability $p_{coh}(t)$ for the collapse and revival for the coherent state given by (27) with $|\beta| = 2$, that is, four initial photons.

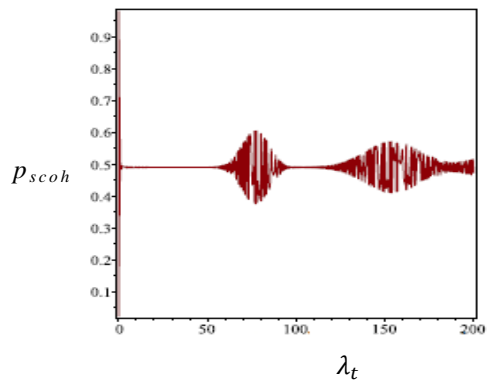

**Fig. 3.** Plot for the probability $p_{scoh}(t)$ for the collapse and revival given in Eq. (26) with $|\alpha| = 0$, $|\beta| = 5$ and $|\zeta| = 0.9$ in (25), that is, with an initial coherent state with twenty five photons.

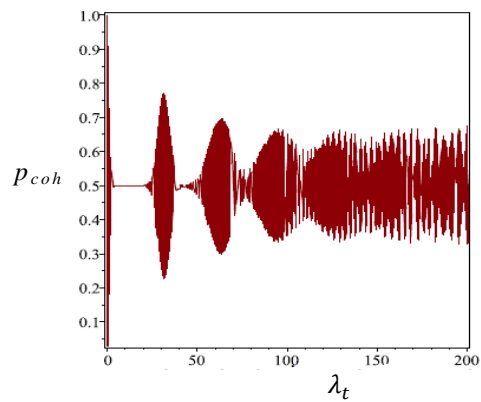

**Fig. 4.** Plot for the probability $p_{coh}(t)$ for the collapse and revival for the coherent state given by (27) with $|\beta| = 5$, that is, twenty five initial photons.

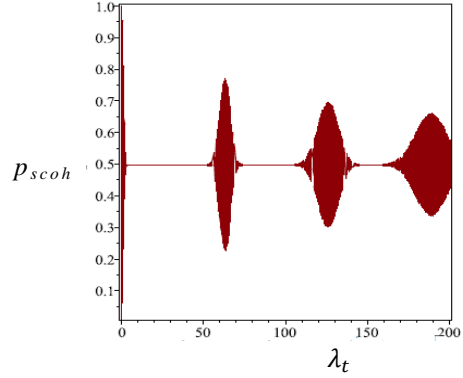

**Fig. 5.** Plot for the probability $p_{scoh}(t)$ for the collapse and revival given in Eq. (26) with $|\alpha| = 15$, $|\beta| = 5$ and $|\zeta| = 0$ in (25), that is, with an initial coherent state with twenty five photons.

It is clear from (17) that in the limit of large values of $r$ and for $\varphi = \pi$, the Hamiltonian (17) for $|\alpha| = 0$ reduces to the original Hamiltonian for the JCM before requiring the rotating-wave approximation [13] with the coupling constant renormalized to $\lambda \to \lambda e^r/2$. In order to numerically study the behavior for large values of $r$, we have plotted in Fig. 6 the behavior of the collapse and revival for $|\alpha| = 0$, $|\beta| = 1$, and $|\zeta| = 2.3$. Unfortunately, the double series in (25) are slowly convergent and so we are limited to numerical evaluations for small values of $r$.

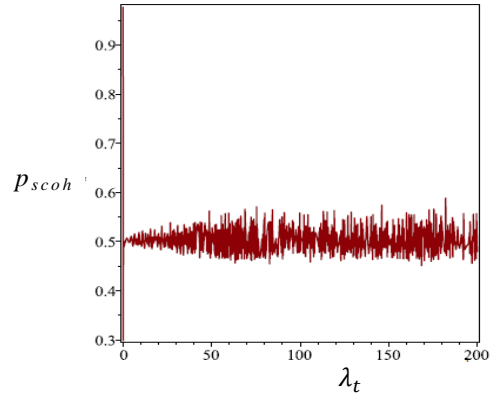

**Fig. 6.** Plot for the probability $p_{scoh}(t)$ for the collapse and revival given in Eq. (26) $|\alpha| = 0$, $|\beta| = 1$ and $|\zeta| = 2.3$ in (25), that is, with an initial coherent state with one photon.

The JCM without the rotating-wave approximations give rise to the following differential equations, viz., Eq. (18) with $\Delta = 0$ and $r = \ln(2)$,

$$\ddot{c}_{1,n}(t) = -\lambda^2(2n+1)c_{1,n}(t) + \lambda^2\sqrt{(n+1)(n+2)}c_{1,n+2}(t) + \lambda^2\sqrt{n(n-1)}c_{1,n-2}(t)$$
$$\ddot{c}_{2,n}(t) = -\lambda^2(2n+1)c_{2,n}(t) + \lambda^2\sqrt{(n+1)(n+2)}c_{2,n+2}(t) + \lambda^2\sqrt{n(n-1)}c_{2,n-2}(t). \qquad (28)$$

The dynamics in this case gives rise to a second-order, linear differential equation consisting of a three-term recursive relation, rather than the non-recursive equations that results in the JCM with the rotating-wave approximation. We have not analytically solved the set (28) since they give rise to an infinitely set of coupled equations. However, we believe the solutions will be sinusoidal functions with frequency $\lambda e^r/2$ and so in the limit of large $r$ the probability $P(t)$ approaches 1/2 for $t > 0$, by the Riemann-Lebesgue lemma, while remaining equal to one at $t = 0$. This behavior is already partially evident in Fig. 6.

## 6. Conclusions

We have generated a new photon Hamiltonian by replacing the creation and annihilation operators for photons in the original Jaynes-Cummings model (JCM) with the creation and annihilation operators of squeezed coherent photons, respectively. This generalized Hamiltonian allows us to study the effects of the terms discard in the JCM when making the rotating-wave approximation. In addition, there are new atom exchange interactions that do not appear in the JCM, which we are able to study the effects that these new additional interactions have on the collapse and revival of an initial coherent state of photons.